\documentclass[manuscript]{aastex}
\usepackage{rotating,graphicx}
\usepackage{amsmath}
\usepackage{cases}

\slugcomment{}

\shorttitle{The Counter-streaming Particle Beams Revisited}
\shortauthors{He}

\begin{document}

\title{Perpendicular Diffusion in the Transport of Solar Energetic Particles from Unconnected Sources: The Counter-streaming Particle Beams Revisited \\}

\author{H.-Q. He\altaffilmark{1,2,3}}

\altaffiltext{1}{Institut f\"{u}r Theoretische Physik, Lehrstuhl IV:
Weltraum- und Astrophysik, Ruhr-Universit\"{a}t Bochum, D-44780
Bochum, Germany; hqhe@mail.iggcas.ac.cn}

\altaffiltext{2}{Key Laboratory of Earth and Planetary Physics,
Institute of Geology and Geophysics, Chinese Academy of Sciences,
Beijing 100029, China}

\altaffiltext{3}{CAS Key Laboratory of Geospace Environment,
Department of Geophysics and Planetary Sciences, University of
Science and Technology of China, Hefei, Anhui 230026, China}

\begin{abstract}
In some solar energetic particle (SEP) events, a counter-streaming
particle beam with a deep depression of flux at $\sim90^{\circ}$
pitch angle during the beginning phase is observed. Two different
interpretations exist in the community to explain this interesting
phenomenon. One explanation invokes the hypothesis of an outer
reflecting boundary or a magnetic mirror beyond the observer. The
other one considers the effect of the perpendicular diffusion on the
transport process of SEPs in the interplanetary space. In this work,
we revisit the problem of the counter-streaming particle beams
observed in SEP events and discuss the possible mechanisms
responsible for the formation of this phenomenon. We clarify some
results in previous works.
\end{abstract}


\clearpage

\section{Introduction}
Solar energetic particles (SEPs) are charged particles of up to GeV
energies occasionally emitted by the Sun. Generally speaking, SEPs
are associated with solar flares or/and coronal mass ejections
(CMEs), although the relative roles of flares and CME-driven shocks
in producing high-energy particles are not completely understood.
Theoretically, SEPs observed in the interplanetary space provide
fundamental information regarding the acceleration and transport
process of charged particles. Therefore, the subject has become a
focus of space physics, astrophysics, and plasma physics.
Furthermore, achieving a better understanding of the transport and
acceleration of charged particles is essential for space weather
research regarding SEP events.

The diffusion mechanism of SEPs transporting in interplanetary space
consists of two components, namely, parallel diffusion along and
perpendicular diffusion across the mean magnetic field. The parallel
diffusion plays an obviously important role in the propagation of
SEPs, so it has been extensively investigated. The perpendicular
diffusion, however, was ignored for quite a long time in previous
studies of SEP transport in interplanetary space. Recently, the
importance of perpendicular diffusion has been gradually and
increasingly realized by the numerical modeling community
\citep[e.g.,][]{Zhang2009,He2011,Giacalone2012,Droge2014,He2015,Strauss2015}
and the observational community
\citep[e.g.,][]{Zhang2003,He2013,Dresing2012,Dresing2014} of SEP
transport and distribution. The effective perpendicular diffusion
and its important effects on the transport process of SEPs were seen
in these numerical or observational studies.

There are some interesting phenomena in the research field of SEP
events, such as SEP reservoirs, east-west longitudinally asymmetric
distribution, and counter-streaming particle beams. The SEP
reservoir refers to the phenomenon where the particle intensities
evolve similarly in time with nearly the same decay rate during the
late phase of the SEP events. The SEP reservoirs are observed by
spacecraft at both low and high heliolatitudes, and by spacecraft at
very different heliolongitudes and radial distances. In addition,
the energetic particle reservoirs are observed during both isolated
SEP events and a sequence of SEP events. The SEP reservoirs are
detected not only in proton data, but also in electron and heavy-ion
data. There exist some different viewpoints in the community to
explain the formation of the SEP reservoirs in the heliosphere. One
explanation is that the perpendicular diffusion mechanism
effectively and uniformly distributes the charged particles in
longitude and latitude to form the observed SEP reservoirs
\citep[e.g.,][]{McKibben1972,McKibben2003,Zhang2009,He2011,He2015}.
Another explanation is that there exist outer reflecting boundaries
or diffusion barriers in the interplanetary space, formed by the
plasma disturbances originating from intense solar bursts, to delay
the SEPs from escaping to larger radial distances and uniformly
redistribute them in longitude and latitude
\citep[e.g.,][]{Roelof1992,Reames1997,Tan2009}. However, as
\citet{He2015} pointed out, it is difficult to imagine such an
``overwhelming" reflecting boundary or diffusion barrier covering
all of the longitudes, all of the latitudes, and even all of the
radial distances. For the formation of SEP reservoirs, there is also
a third explanation combining the two aforementioned mechanisms
\citep[e.g.,][]{Lario2003}.

The east-west longitudinally asymmetric distribution of SEPs refers
to the phenomenon that with the same longitude separation between
the solar source and the magnetic footpoint of the observer, the
flux of the SEP event originating from the solar source located at
eastern side of the nominal magnetic footpoint of the observer is
larger than that of the SEP event originating from the source
located at the western side. This phenomenon was found in the
numerical simulations of the multidimensional transport equation of
SEPs \citep{He2011,He2015} and was also proven by the spacecraft
observations \citep{He2013,Dresing2014}. We conclude that the
longitudinally asymmetric distribution of SEPs results from the
east-west azimuthal asymmetry in the geometry of the Parker
interplanetary magnetic field as well as the effects of
perpendicular diffusion on the transport processes of SEPs in the
heliosphere \citep{He2011,He2013,He2015}. Other interpretations for
this phenomenon also exist \citep[e.g.,][]{Lario2014}.

The counter-streaming particle beam is the phenomenon where during
the onset phase of an SEP event, a significant number of particles
could move toward the Sun, while the SEP intensities at
$\sim90^{\circ}$ pitch angle still stay deeply depressed. Two
different mechanisms have been provided to explain this SEP
phenomenon. On the one hand, some authors suggested that the
counter-streaming particle beams observed in the SEP events result
from an outer reflecting boundary or a nearby magnetic mirror
outside of 1 AU \citep[e.g.,][]{Tan2009,Tan2011}. They further
suggested that a counter-streaming particle beam with a deep
depression around $90^{\circ}$ pitch angle during the beginning of
the SEP event is an evidence for the presence of an outer reflecting
boundary. On the other hand, a model calculation of SEP transport
with the effect of perpendicular diffusion in the three-dimensional
interplanetary magnetic field showed that the counter-streaming
particle beams with a deep depression at $\sim90^{\circ}$ pitch
angle can be reproduced without invoking the hypothesis of an outer
reflecting boundary or a magnetic mirror \citep{Qin2011}. Other
transport simulations including the effects of pitch-angle diffusion
and reflecting boundary were also presented to account for the
back-streaming electrons \citep[e.g.,][]{Kartavykh2013}. However,
\citet{Tan2012} argued against the existence of scattering and
perpendicular diffusion experienced by the low energy electrons in
the SEP events and stated that ``the simulation result shown in
Figure 3 of \citet{Qin2011} does not exhibit a particle depression
at $\mu\sim0$ as observed by \citet{Tan2009}, contrary to their
claim". Therefore, it is necessary and important to revisit the
problem of the counter-streaming particle beams and clarify some
confusing results in the previous works.

In this paper, we revisit the phenomenon of the counter-streaming
particle beams observed in the onset phase of SEP events to clarify
some confusing results in the previous works. Specifically, in
Section 2, we present the spacecraft observation of the
counter-streaming particle beam in the 2001 September 24 SEP event.
In Section 3, we present a model calculation of SEP propagation with
perpendicular diffusion in the three-dimensional interplanetary
magnetic field, and fit to the SEP observation in the 2001 September
24 event. In Section 4, we discuss the fitting results and the
possible mechanisms responsible for the phenomenon of the
counter-streaming particle beams. A summary of our results will be
provided in Section 5.

\section{Counter-streaming Particle Beam: 2001 September 24 SEP Event}
The solar flare associated with the 2001 September 24 SEP event
occurred at S16E23, starting at 09:32 UT, reaching optical emission
maximum at 10:38 UT, and ending at 11:09 UT. This SEP event was
observed by the Three-Dimensional Plasma (3DP) instrument onboard
the Wind spacecraft. During the beginning phase of this SEP event,
the so-called counter-streaming particle beam phenomenon was
observed. The red solid lines in Figure \ref{20010924-exact-time}
indicate the $\sim40$ keV electron pitch-angle distributions
observed by the Wind/3DP instrument during the 2001 September 24 SEP
event. The angular distributions of particles are obtained from
sectored measurements of eight directions. The time resolution of
the Wind/3DP data we use in this work is 5 s. The increasing time
labeled in red from the bottom to top panels indicates when the
pitch-angle distribution data were measured by Wind/3DP. The panels
from bottom to top are plotted in time intervals of 15 minutes.
Following previous works related to this SEP event, the X-axis
coordinate in Figure \ref{20010924-exact-time} is set so that the
pitch-angle cosine $\mu=-1$ corresponds to charged particles moving
away from the Sun.

The Wind/3DP observations, denoted by the red solid lines in Figure
\ref{20010924-exact-time}, show that in the very beginning of the
2001 September 24 SEP event, the SEP flux only arises on the right
side, i.e., at $\mu\lesssim-0.2$. It means that all particles are
nearly propagating away from the Sun. After 30 minutes, however,
electrons with $\mu\gtrsim0.7$ start to appear and increase
gradually. Interestingly, during this evolution process, the
electrons with $\mu\sim0$ were still absent or had very low
intensity. Therefore, the so-called counter-streaming particle beam
phenomenon was observed in this SEP event. \citet{Tan2009} proposed
that there is no scattering of incident electrons because the flux
near $90^{\circ}$ pitch-angle (corresponding to $\mu\sim0$) is
nearly $0$, and further suggested that the electron flux increases
at $\mu\gtrsim0.7$ are mainly formed by reflected electrons from a
stronger magnetic field configuration beyond the location of the
observer, i.e., the Wind spacecraft at 1 AU. \citet{Tan2009} claimed
that the presence of a counter-streaming particle beam with an
intensity depression of electrons at $\sim90^{\circ}$ pitch angle
($\mu\sim0$) could be a strong evidence for the existence of a
nearby outer reflecting boundary of SEPs.

In the next section, we will present a numerical model of SEP
propagation with perpendicular diffusion in the three-dimensional
interplanetary magnetic field. The simulation results will be used
to explain the observed counter-streaming particle beam in the 2001
September 24 SEP event.

\section{Numerical Simulations of the 2001 September 24 SEP Event}
\subsection{Numerical Model}
The multidimensional Fokker-Planck focused transport equation that
governs the gyrophase-averaged distribution function
$f(\textbf{x},\mu,p,t)$ of charged particles can be written as
\citep[e.g.,][]{Schlickeiser2002,Zhang2009,He2011,He2015,Droge2010,Droge2014}
\begin{eqnarray}
{}&&\frac{\partial f}{\partial t}+\mu v\frac{\partial f}{\partial
z}+{\bf V}^{sw}\cdot\nabla f+\frac{dp}{dt}\frac{\partial f}{\partial
p}+\frac{d\mu}{dt}\frac{\partial f}{\partial \mu}  \nonumber\\
{}&&-\frac{\partial}{\partial\mu}\left(D_{\mu\mu}\frac{\partial
f}{\partial \mu}\right)-\frac{\partial}{\partial
x}\bigg(\kappa_{xx}\frac{\partial f}{\partial x}\bigg)
-\frac{\partial}{\partial y}\left(\kappa_{yy}\frac{\partial
f}{\partial y}\right)=Q({\bf x},p,t),  \label{transport-equation}
\end{eqnarray}
where $\textbf{x}$ is the particle's position, $z$ is the coordinate
along the magnetic field spiral, $p$ is the particle's momentum,
$\mu$ is the pitch-angle cosine of the particle, $t$ is time, $v$ is
the particle speed, $\textbf{V}^{sw}$ is the solar wind speed,
$\kappa_{xx}$ and $\kappa_{yy}$ are the perpendicular diffusion
coefficients, and $Q$ is the source term. The term $dp/dt$
represents the adiabatic cooling effect and can be written as
\citep[e.g.,][]{Skilling1971}
\begin{equation}
\frac{dp}{dt}=-p\left[\frac{1-\mu^2}{2}\left(\frac{\partial
V^{sw}_x}{\partial x}+\frac{\partial V^{sw}_y}{\partial
y}\right)+\mu^2\frac{\partial V^{sw}_z}{\partial z}\right].
\label{adiabatic-cooling}
\end{equation}
The term $d\mu/dt$, including the effects of magnetic focusing and
the divergence of solar wind flows, can be written as
\citep[e.g.,][]{Roelof1969,Isenberg1997,Kota1997}
\begin{eqnarray}
\frac{d\mu}{dt}&=&\frac{1-\mu^2}{2}\left[-\frac{v}{B}\frac{\partial
B}{\partial z}+\mu \left(\frac{\partial V^{sw}_x}{\partial
x}+\frac{\partial V^{sw}_y}{\partial y}-2\frac{\partial
V^{sw}_z}{\partial z}\right)\right]     \nonumber\\
{}&=&\frac{1-\mu^2}{2}\left[\frac{v}{L}+\mu\left(\frac{\partial
V^{sw}_x}{\partial x}+\frac{\partial V^{sw}_y}{\partial
y}-2\frac{\partial V^{sw}_z}{\partial z}\right)\right],
\label{magnetic-focusing}
\end{eqnarray}
where $B$ is the average interplanetary magnetic field, and the
magnetic focusing length $L$ is defined by
$L=(\textbf{z}\cdot\bigtriangledown \ln B)^{-1}$.

We use the form of the pitch-angle diffusion coefficient as
\citep[e.g.,][]{Beeck1986,He2011,He2015}
\begin{equation}
D_{\mu\mu}^{r}=D_{\mu\mu}/\cos^{2}\psi=D_{0}vR^{-1/3}\left(|\mu|^{q-1}+h\right)(1-\mu^{2}),
\label{diffusion-coefficient}
\end{equation}
where $D_{0}$ is a constant indicating the magnetic turbulence
strength and $R$ is the particle rigidity. The constant $h$ is
needed to model the particles' scattering ability through $\mu=0$
($90^{\circ}$ pitch angle). In order to simulate the nonlinear
effect to cause large $D_{\mu\mu}$ at $\mu=0$, we use a relatively
large value of $h=0.2$. The constant $q$ is related to the power
spectrum of magnetic turbulence in the inertial range, chosen to be
$5/3$ in our numerical model.

We employ the time-backward Markov stochastic process method to
numerically solve the five-dimensional Fokker-Planck transport
equation (\ref{transport-equation}). After the operation with the
Markov stochastic process method, we can obtain five time-backward
stochastic differential equations, which are recast from the
Fokker-Planck transport equation (\ref{transport-equation}), as
follows \citep[e.g.,][]{Zhang2009,He2011,He2015}:
\begin{eqnarray}
dX &=& \sqrt{2\kappa_{xx}}dW_{x}(s)-V_{x}^{sw}ds  \nonumber\\
dY &=& \sqrt{2\kappa_{yy}}dW_{y}(s)-V_{y}^{sw}ds  \nonumber\\
dZ &=& -(\mu V+V_{z}^{sw})ds  \nonumber\\
d\mu &=& \sqrt{2D_{\mu\mu}}dW_{\mu}(s)  \nonumber\\
{}&& -\frac{1-\mu^{2}}{2}\left[\frac{V}{L}+\mu\left(\frac{\partial
V_{x}^{sw}}{\partial x}+\frac{\partial V_{y}^{sw}}{\partial
y}-2\frac{\partial V_{z}^{sw}}{\partial z}\right)\right]ds  \nonumber\\
{}&& +\left(\frac{\partial D_{\mu\mu}}{\partial
\mu}+\frac{2D_{\mu\mu}}{M+\mu}\right)ds  \nonumber\\
dP &=& P\left[\frac{1-\mu^{2}}{2}\left(\frac{\partial
V_{x}^{sw}}{\partial x}+\frac{\partial V_{y}^{sw}}{\partial
y}\right)+\mu^{2}\frac{\partial V_{z}^{sw}}{\partial z}\right]ds,
\label{eq:stochastic-process}
\end{eqnarray}
where $(X,Y,Z)$ is the pseudo-position, $V$ is the pseudo-speed, $P$
is the pseudo-momentum, and $W_{x}(t)$, $W_{y}(t)$, and $W_{\mu}(t)$
are Wiener processes.

In Equation (\ref{transport-equation}), the source term $Q$ is
assumed to be the following form \citep[e.g.,][]{Reid1964,He2011}
\begin{equation}
Q(r\leqslant0.01AU,E_{k},\theta,\phi,t)=\frac{C}{t}\frac{E_{k}^{-\gamma}}{p^{2}}
\exp\left(-\frac{\tau_{c}}{t}-\frac{t}{\tau_{L}}\right)\xi(\theta,\phi),
\label{source}
\end{equation}
where $\gamma$ is the power-law spectrum index of source particles,
set to be $3$, $\tau_{c}$ and $\tau_{L}$ are the time constants
indicating the rise and decay timescales of the particle release
profile in the solar corona, which are set to be $\tau_{c}=3.30$ hr
and $\tau_{L}=6.74$ hr in this work, respectively, and
$\xi(\theta,\phi)$ is a function describing the latitudinal and
longitudinal distribution of SEP injection strength in solar
sources. In this work, an SEP source with limited coverage in
latitude and longitude is used. In the numerical simulations, we
need to use an outer absorptive boundary of pseudoparticles, which
is set at $r=50$ AU in this work. According to the Wind spacecraft
measurements during the 2001 September 24 SEP event, in the
simulations we use a solar wind speed $V^{sw}=450~km~s^{-1}$.

The numerical modeling results were usually normalized to compare
with the observation data of the SEP events, which will be done in
the following. In each figure of simulation results that follows,
the normalization factor is the same for plotting all the panels of
that figure. We also note that in each figure, the normalization
factor for plotting the simulation results with perpendicular
diffusion is different from that for plotting the simulation results
without perpendicular diffusion, since the particle intensity of the
latter case is much larger than that of the former case, due to a
uniform SEP source used in the latter case.

\subsection{Simulation Results}
\begin{figure}
 \epsscale{1.0}
 \plotone{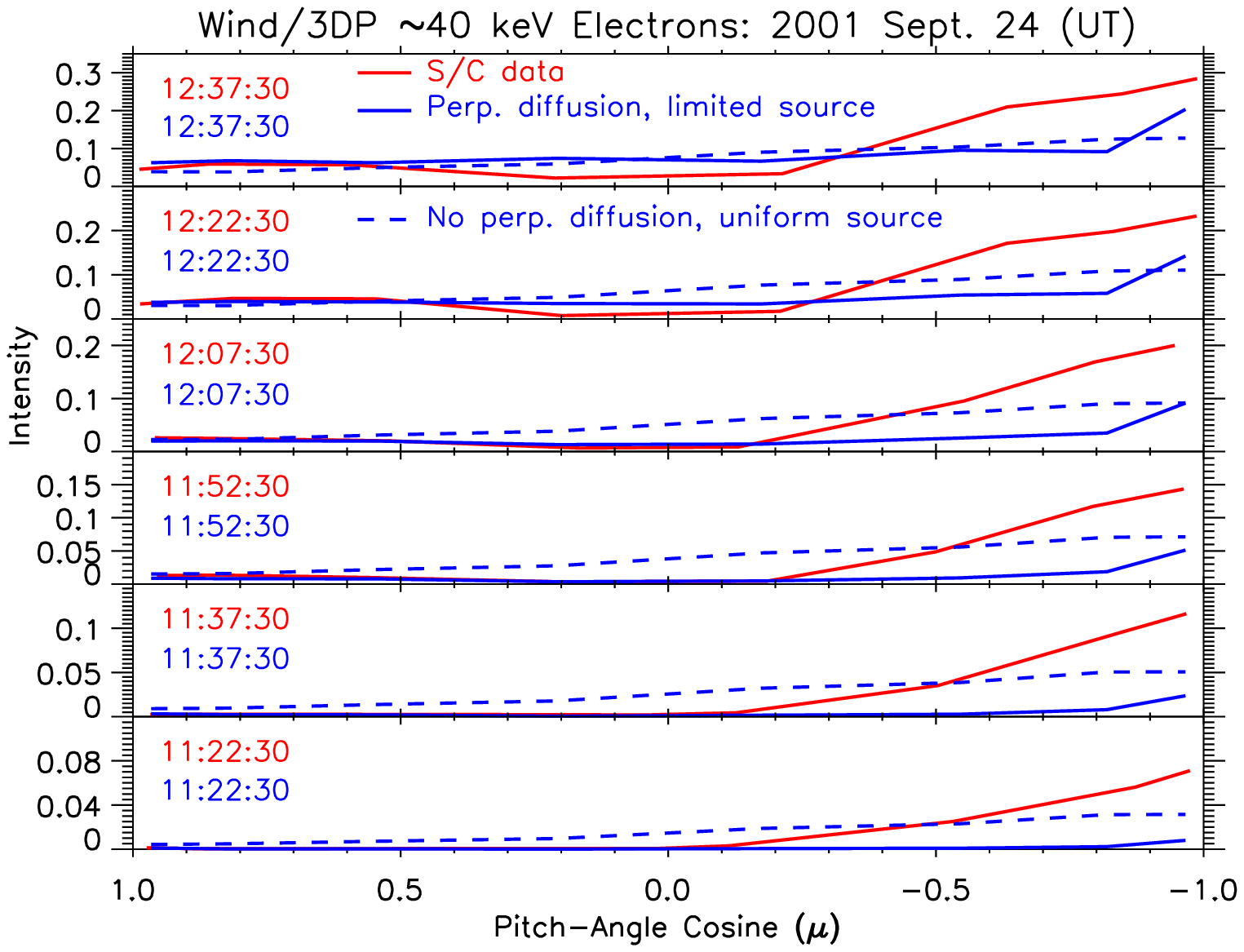}
 \caption{Temporal evolution (from bottom to top panels) of pitch-angle distributions of
 $\sim40$ keV electrons during the beginning stage of the 2001 September 24 SEP event.
 In each panel, the red solid line indicates the spacecraft data; the blue solid and
 dashed lines indicate the simulation results with and without perpendicular diffusion,
 respectively. The simulation time (blue) in each panel is set to the exact same time as
 the observation time (red). \label{20010924-exact-time}}
\end{figure}

In this case study, we set the radial mean free path to be constant
with the value $\lambda_{r}=0.15$ AU (corresponding to the parallel
mean free path $\lambda_{\parallel}=0.34$ AU at 1 AU). In the
simulations with perpendicular diffusion, the perpendicular mean
free paths are set to the constant value
$\lambda_{x}=\lambda_{y}=0.03$ AU for 40 keV electrons. The blue
solid lines in Figure \ref{20010924-exact-time} present the
simulation results of the 2001 September 24 SEP event. The
increasing time labeled in blue in the bottom to top panels
indicates when the particle pitch-angle distributions occur in the
numerical simulations. As we can see, the simulation time in each
panel is set the exact same as the observation time. The evolution
behavior of the particles' pitch-angle distribution in the
simulation scenario with perpendicular diffusion is similar to the
Wind/3DP observation of the 2001 September 24 SEP event. In the
beginning, the observer at 1 AU only detected the energetic
electrons moving away from the Sun, i.e., streaming in the
anti-sunward direction. Later at around 11:52:30 UT, the electrons
with pitch-angle cosine $\mu\gtrsim0.5$ began to appear. After that,
the electron intensities in both the anti-sunward and sunward
directions gradually increased with time, while during this period
the electron flux at $\mu\sim0$ was still depressed. In the top two
panels, the simulated electron flux at $\mu\sim0$ gradually
increased. At late stage of the 2001 September 24 SEP event, the
Wind/3DP observation also showed that the electron intensity at
$\mu\sim0$ began to appear. The blue dashed lines in Figure
\ref{20010924-exact-time} indicate the simulation results without
perpendicular diffusion, but with a uniform SEP source. One can see
that in this scenario, the electron intensity in each panel
monotonically decreases with increasing pitch-angle cosine $\mu$,
without any intensity depression at $\mu\sim0$. Therefore, the
direct comparison between the simulation results obtained with and
without perpendicular diffusion suggests that the perpendicular
diffusion can cause the so-called counter-steaming particle beams
with a deep intensity depression at $\mu\sim0$.

As we know, in the interplanetary space, the physical circumstances
and conditions for SEP diffusion and propagation are very
complicated and dynamic. In our simulations of SEP propagation, we
used some necessary assumptions for simplification in the numerical
modeling. Therefore, it is difficult to exactly simulate the entire
temporal evolution of SEP events at a given position in the
interplanetary space. In Figure \ref{20010924-exact-time}, in order
to match the timing of the 2001 September 24 SEP event, we set the
simulation time in each panel the exact same as the observation
time. Therefore, the fitting of the particle pitch-angle
distribution is not very perfect. However, the basic evolution
features of the 40 keV electrons in the 2001 September 24 SEP event
have been successfully reproduced in our numerical simulations based
on the five-dimensional Fokker-Planck transport equation.

\begin{figure}
 \epsscale{1.0}
 \plotone{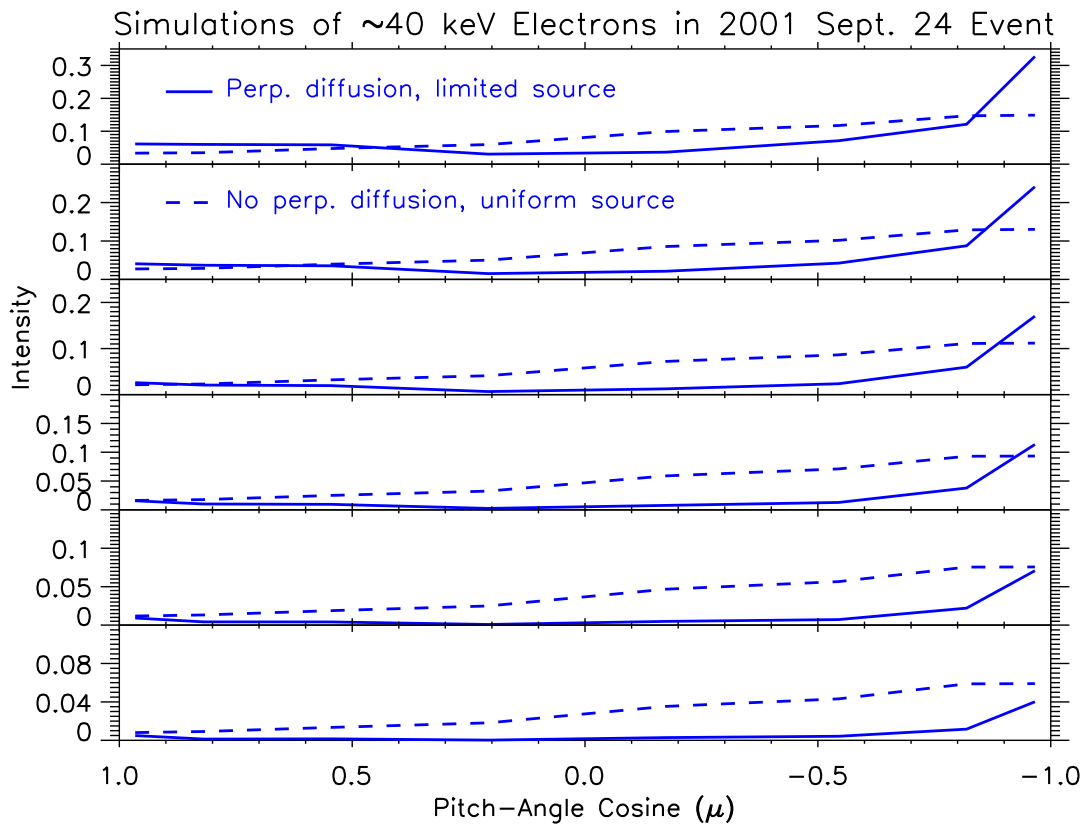}
 \caption{The simulation results are plotted in a rough-timing manner. The blue solid and
 dashed curves indicate the simulation results with and without perpendicular diffusion, respectively.
 The counter-streaming particle beam with deep depression at $\sim90^{\circ}$ pitch angle is qualitatively
 reproduced in the simulation with perpendicular diffusion. \label{20010924-rough-time}}
\end{figure}

To better fit the particles' pitch-angle distribution, we replot the
simulation results with perpendicular diffusion in Figure
\ref{20010924-rough-time} without exactly matching the timing of the
SEP event. The blue solid curves in Figure \ref{20010924-rough-time}
show the numerical simulation results with perpendicular diffusion,
plotted in a rough-timing manner. We note that the normalization
factors for plotting the simulation results in Figure
\ref{20010924-rough-time} are different from those in Figure
\ref{20010924-exact-time}. As one can see, the simulation results
with perpendicular diffusion show excellent agreement with the
evolution of the electron pitch-angle distributions (shown with red
solid lines in Figure \ref{20010924-exact-time}) observed by the
Wind/3DP instrument during the beginning of the 2001 September 24
SEP event. In this rough-timing fashion, a much better fitting of
the simulation with perpendicular diffusion to the observation data
is obvious to see. In the very beginning of the SEP event, only the
energetic electrons moving away from the Sun were observed at 1 AU.
Later, the energetic electrons at $\mu\gtrsim0.5$ began to appear.
With increasing time, the intensities of electrons both moving away
from and toward the Sun gradually increased. However, the electrons
with $\mu\sim0$ were still absent during this period. Therefore, the
counter-streaming particle beam with a deep depression at
$\sim90^{\circ}$ pitch angle was evidently reproduced in the
numerical simulation with perpendicular diffusion. The blue dashed
curves in Figure \ref{20010924-rough-time} present the simulation
results without perpendicular diffusion. As shown in Figure
\ref{20010924-exact-time}, during the SEP event onset the simulation
without perpendicular diffusion in Figure \ref{20010924-rough-time}
does not display any observational features of the so-called
counter-streaming particle beam with deep intensity depression at
$\mu\sim0$. The strong comparison between the simulation results
with and without perpendicular diffusion suggests that the
counter-steaming particle beams can be caused by the perpendicular
diffusion of SEPs.

\section{Discussion on the Possible Mechanisms Responsible for Counter-streaming Particle Beams}
The work \citet{Tan2009} discussed the phenomenon of the
counter-streaming particle beams observed in SEP events and proposed
that an outer reflecting boundary or a nearby magnetic mirror
outside of 1 AU causes this SEP phenomenon. Furthermore, they
suggested that a counter-streaming particle beam with a deep
intensity depression around the $90^{\circ}$ pitch angle during the
beginning stage of the SEP event is an evidence for the presence of
an outer reflecting boundary. In some specific circumstances, an
outer reflecting boundary or a magnetic mirror could be formed by
large-scale plasma disturbances launched during periods of intense
solar activity. The enhanced magnitude of the interplanetary
magnetic field formed at these reflecting boundaries could influence
the transport processes of charged particles, even change their
transport directions in the interplanetary space, e.g., from the
anti-sunward to sunward directions. If the number of particles
streaming toward the Sun after being reflected from a reflecting
boundary is quite considerable, and the particles crossing the
magnetic field lines are very scarce, then the so-called
counter-streaming particle beam with a deep depression of flux at
$\sim90^{\circ}$ pitch angle could form. However, it requires that
the configuration of the disturbed interplanetary magnetic field is
exquisite enough. In addition, so far no explicit or quantitative
mechanism of the outer reflecting boundaries has been specified in
the community. Therefore, to present a mathematically tractable
description should be the critical task for the mechanism of
reflecting boundary.

\begin{figure}
 \epsscale{1.0}
 \plotone{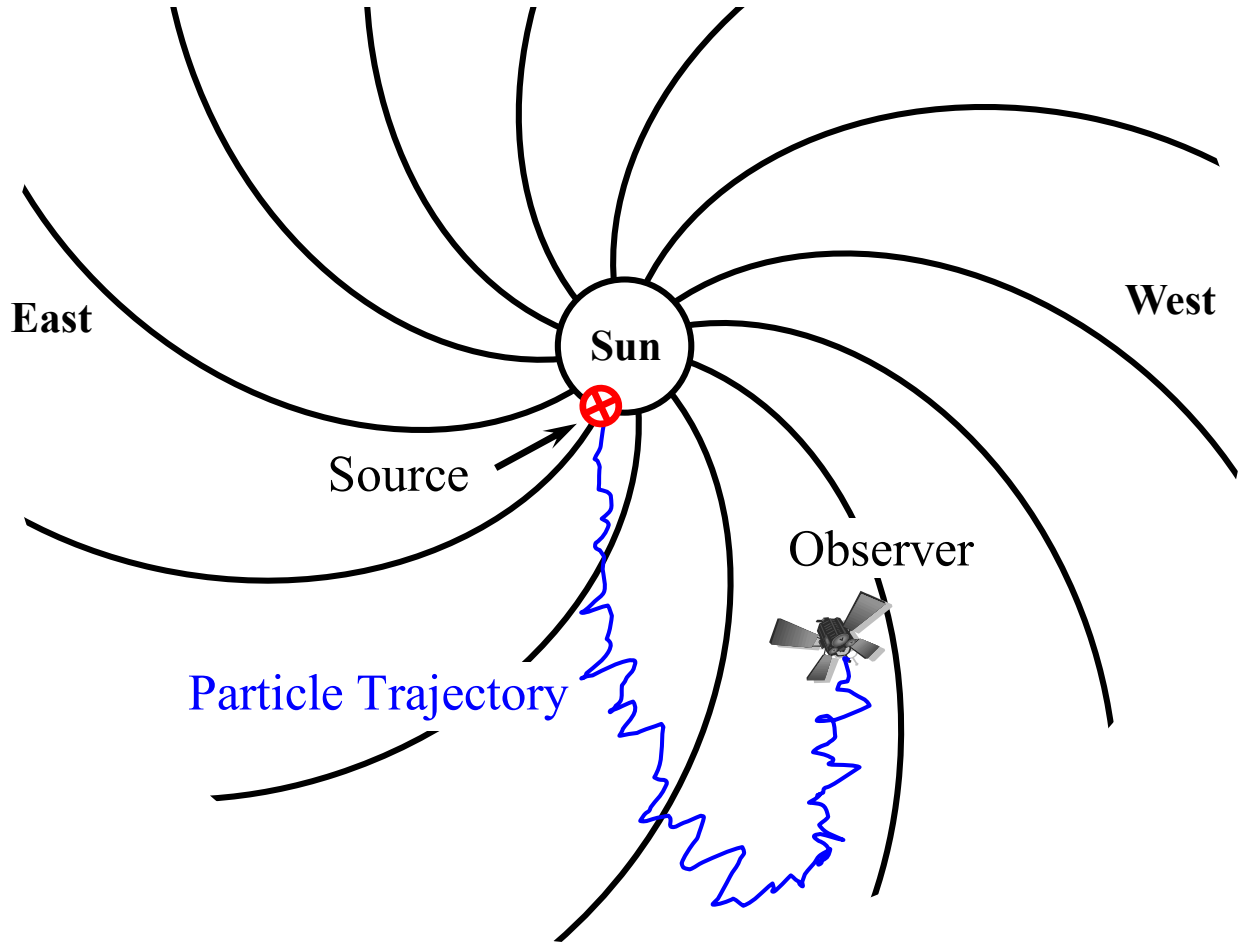}
 \caption{Illustration of the formation of counter-streaming particle beams in
the interplanetary magnetic field. \label{Illustration}}
\end{figure}

Generally, the propagation of SEPs in the interplanetary magnetic
field mainly consists of the along-field and the cross-field
processes, known as the parallel and the perpendicular components,
respectively. In this work, we provide an alternative mechanism for
the formation of counter-streaming particle beams. In our
explanation, the perpendicular diffusion plays a significant role in
the formation of counter-streaming particle beams. Our explanation
is based on the numerical modeling of the multidimensional
Fokker-Planck transport equation. In our modeling, we use a
Parker-type interplanetary magnetic field and a constant solar wind
speed $V^{sw}=450~km~s^{-1}$. Accordingly, we can easily deduce that
during the 2001 September 24 SEP event, the nominal magnetic
footpoint of the observer (i.e., the Wind spacecraft) was at
$\sim50.4^{\circ}$ W in heliographic longitude and $\sim7.0^{\circ}$
N in heliographic latitude. Therefore, the separation between the
nominal magnetic footpoint of the Wind spacecraft and the solar
flare location (S16E23) was $\sim73.4^{\circ}$ in longitude and
$\sim23.0^{\circ}$ in latitude. In addition to parallel diffusion
along the mean magnetic field, the SEPs will experience
perpendicular diffusion and cross the interplanetary magnetic field.
Therefore, the SEPs originating from acceleration regions with
limited coverage can be observed by distant spacecraft with large
longitudinal or/and latitudinal separations. In our simulation with
perpendicular diffusion, we use $\lambda_r=0.15$ AU and
$\lambda_x=\lambda_y=0.03$ AU. Near the Sun, the magnetic focusing
effect plays an important role in the transport of SEPs; while at
larger radial distances, the significance of the adiabatic focusing
is largely reduced
\citep[e.g.,][]{Schlickeiser2008,He2012a,He-Schlickeiser2015,Shalchi2013,Tautz2014}.
Due to the relatively large parallel mean free path used in the
modeling, quite a number of energetic electrons will transport
approximately along the magnetic field lines and reach radial
distances larger than 1 AU before being scattered back. Meanwhile,
as a result of perpendicular diffusion, the energetic electrons will
move in the perpendicular direction to cross the magnetic field
lines, particularly at larger radial distances. These cross-field
electrons can move onto the field line connecting the observer whose
magnetic footpoint is tens of degrees in longitude and latitude away
from the SEP source. Afterwards, some of these electrons will be
scattered back there and move toward the Sun, i.e., in the sunward
direction. The observer at $\sim1$ AU can detect two different types
of particles: some particles crossed the field lines at smaller
radial distances $r\lesssim 1$ AU; while other particles crossed the
field lines at larger radial distances $r\gtrsim1$ AU. The former
particles are observed as moving in the anti-sunward direction;
while the latter particles are detected as streaming in the sunward
direction. Therefore, in the beginning of the SEP event, a
counter-streaming particle beam with a deep depression of intensity
at $\sim90^{\circ}$ pitch-angle is formed. After that, particles
with intense scattering begin to appear at the observer, so the
intensity at $\mu\sim0$ begins to increase. The physical scenario of
the formation of counter-streaming particle beams via perpendicular
diffusion is sketched in Figure \ref{Illustration}.

We note that in this work, we use some simplifications in the
modeling of SEP transport. For instance, we use a Parker
interplanetary magnetic field and a constant solar wind speed
$V^{sw}=450~km~s^{-1}$. For the diffusion coefficients, we use a
constant radial mean free path $\lambda_r=0.15$ AU and constant
perpendicular mean free paths $\lambda_x=\lambda_y=0.03$ AU.
Additionally, we adopt the SEP source function as in Equation
(\ref{source}). Due to the very dynamic interplanetary conditions,
the realistic transport parameters and expressions should be more
complicated. For example, in more accurate modeling of SEP
transport, the pitch-angle dependence \citep[e.g.,][]{Strauss2015}
and radial dependence \citep[e.g.,][]{He2012b} of the perpendicular
diffusion coefficient should be taken into account. In spite of the
simplified SEP model, the numerical calculation with perpendicular
diffusion reproduces the counter-streaming particle beams with a
deep depression of intensity at $\mu\sim0$ during the onset phase,
similar to what was detected by the Wind spacecraft. Our numerical
investigation suggests that the perpendicular diffusion can be a
possible mechanism for the formation of this SEP phenomenon.
Basically, to observe a counter-streaming particle beam requires
some specific conditions of the SEP event, such as a limited source,
observers disconnected from the source, the diffusion processes
(both parallel and perpendicular) of particles, and appropriate
locations of the observers in the interplanetary magnetic field.

Another possible explanation for the formation of counter-streaming
particle beams is a combination of both the reflecting boundary
mechanism and the perpendicular diffusion mechanism. As pointed out
above, this explanation also requires a computationally tractable
description of the reflecting boundaries.

\section{Summary and Conclusions}
In this work, we revisit the so-called counter-streaming particle
beams observed in the SEP events and clarify some confusing results
in the previous works. Our numerical simulations of SEP propagation
with perpendicular diffusion in the three-dimensional interplanetary
magnetic field reproduce the phenomenon of counter-streaming
particle beams with a deep depression of flux at $\mu\sim0$ during
the onset phase of the SEP events. The comparison between the
simulation results with and without perpendicular diffusion proposes
that the perpendicular diffusion can be a possible mechanism
responsible for the formation of the counter-streaming particle
beams. Without exactly matching the timing of the SEP event, the
simulation results with perpendicular diffusion show excellent
agreement with the observation data. Therefore, the claim by
\citet{Tan2012} that ``the simulation result shown in Figure 3 of
\citet{Qin2011} does not exhibit a particle depression at $\mu\sim0$
as observed by \citet{Tan2009}, contrary to their claim" is
incorrect.

Our simulations with perpendicular diffusion reproduce the so-called
counter-streaming particle beams with a deep depression of intensity
at $\mu\sim0$ during the onset phase without invoking the hypothesis
of a reflecting boundary. It indicates that, at least, the
reflecting boundaries are not the only possible reason causing the
counter-streaming particle beams with a deep depression at
$\mu\sim0$ during the onset phase. Therefore, the counter-streaming
particle beams with a deep depression at $\mu\sim0$ during the onset
phase cannot be used as strong evidence for the presence of the
so-called outer reflecting boundaries in the interplanetary space.


\acknowledgments This work was supported in part by grant Schl
201/29-1 of the Deutsche Forschungsgemeinschaft, the National
Natural Science Foundation of China under grants 41204130, 41474154,
41321003, and 41131066, the National Important Basic Research
Project under grant 2011CB811405, the Chinese Academy of Sciences
under grant KZZD-EW-01-2, and the Open Research Program from Key
Laboratory of Geospace Environment, Chinese Academy of Sciences.
H.-Q. He gratefully acknowledges the support of the International
Postdoctoral Exchange Fellowship Program of China under grant
20130023. We benefited from discussions at the team meeting
``Exploration of the inner Heliosphere - what we have learned from
Helios and what we want to study with Solar Orbiter", supported by
the International Space Science Institute (ISSI) in Bern,
Switzerland. We benefited from the Wind/3DP data provided by Space
Physics Research Group (SPRG), Space Sciences Laboratory (SSL),
University of California at Berkeley
(http://sprg.ssl.berkeley.edu/wind3dp/).


\clearpage


\clearpage

\clearpage

\clearpage


\end{document}